\documentclass[9pt,twocolumn,twoside]{pnas-new}
\usepackage{color} 

\templatetype{pnasresearcharticle} 

\title{Switch between critical percolation modes in city traffic dynamics}

\author[a,b]{Guanwen Zeng}
\author[a,b,1]{Daqing Li}
\author[c]{Shengmin Guo}
\author[d]{Liang Gao}
\author[d,1]{Ziyou Gao}
\author[e,f,1]{H. Eugene Stanley}
\author[g]{Shlomo Havlin}

\affil[a]{School of Reliability and Systems Engineering, Beihang University, Beijing 100191, China.}
\affil[b]{Science and Technology on Reliability and Environmental Engineering Laboratory, Beijing 100191, China.}
\affil[c]{State Key Laboratory of Software Development Environment, Beihang University, Beijing 100191, China.}
\affil[d]{School of Traffic and Transportation, Beijing Jiaotong University, Beijing 100044, China.}
\affil[e]{Center for Polymer Studies, Boston University, Boston, MA 02215.}
\affil[f]{Physics Department, Boston University, Boston, MA 02215.}
\affil[g]{Department of Physics, Bar-Ilan University, Ramat Gan 52900, Israel.}

\correspondingauthor{\textsuperscript{1}To whom correspondence should be addressed. \\*E-mail: daqingl@buaa.edu.cn, zygao@bjtu.edu.cn or hes@bu.edu}


\begin{abstract}
Percolation transition is widely observed in networks ranging from biology to engineering. While much attention has been paid to network topologies, studies rarely focus on critical percolation phenomena driven by network dynamics. Using extensive real data, we study the critical percolation properties in city traffic dynamics. Our results suggest that two modes of different critical percolation behaviors are switching in the same network topology under different traffic dynamics. One mode of city traffic (during nonrush hours or days off) has similar critical percolation characteristics as small world networks, while the other mode (during rush hours on working days) tends to behave as a 2D lattice. This switching behavior can be understood by the fact that the high-speed urban roads during nonrush hours or days off (that are congested during rush hours) represent effective long-range connections, like in small world networks. Our results might be useful for understanding and improving traffic resilience.
\end{abstract}

\begin{document}

\verticaladjustment{-2pt}

\maketitle
\thispagestyle{firststyle}
\ifthenelse{\boolean{shortarticle}}{\ifthenelse{\boolean{singlecolumn}}{\abscontentformatted}{\abscontent}}{}

\dropcap{C}ritical phenomena of complex networks ranging from biology to engineering have attracted much attention \cite{RN11}. Studies on percolation \cite{RN1, RN2}, epidemic spreading \cite{RN4, RN6}, the Ising model \cite{RN8} etc. have uncovered the state transition process of different complex systems. As a typical dynamic complex system, traffic is frequently observed to have transitions between free flow and congestion states \cite{RN17, RN18, RN19, RN20}, where the critical transition point marks the possible balance between traffic supply and demand \cite{RN33}. With the rising traffic demand during urban agglomeration, the inescapable congestion will cost not only huge economic losses but also, generate pollutions and even psychological stress from road rage. Accordingly, many models have been applied to analyze the dynamic properties of traffic systems at macro \cite{RN21, RN22, RN23, RN25} or micro \cite{RN26, RN27, RN57} levels.

Different transition processes can be characterized by a unified approach if they have the same critical exponents, a characterization called "universality class" \cite{RN12, RN13}. These critical exponents determine the system behavior near the critical point of transition, which may help us to understand the system robustness and design a resilient system \cite{RN51, RN58, RN10}. For a given instance, urban traffic is organized as a network of local flows, which will become disintegrated under increasing perturbations. The system behavior near the critical point of this disintegration process determines the global operating efficiency of urban traffic and also, the possible mitigation strategies above the transition point. This urges us to explore the transition type of urban traffic, especially the unique way of its disintegration.

Instead of a 1D highway, urban traffic is the result of 2D spatial organization of functional roads, which can be described as a percolation process. The giant component of traffic percolation spanning over the whole-city traffic network will disintegrate into small clusters when only considering functional roads above a critical velocity threshold. Since a percolation transition can be characterized by critical exponents, a fundamental question can be raised: whether the percolation disintegration of city traffic during different traffic periods, such as rush hours and nonrush hours, belongs to the same or different universality classes. While previous studies have focused on effects of static network topology on critical exponents, the critical percolation behaviors of dynamic traffic network, to the best of our knowledge, have not been studied. The answer to the above question can help us to better understand the relation between formation and dissipation of traffic congestion. As we show here, our results indicate that the two modes are found to belong to different universality classes characterized by different critical exponents.

\section*{Results}

\begin{figure*}
\centering
\includegraphics[width=17.8cm]{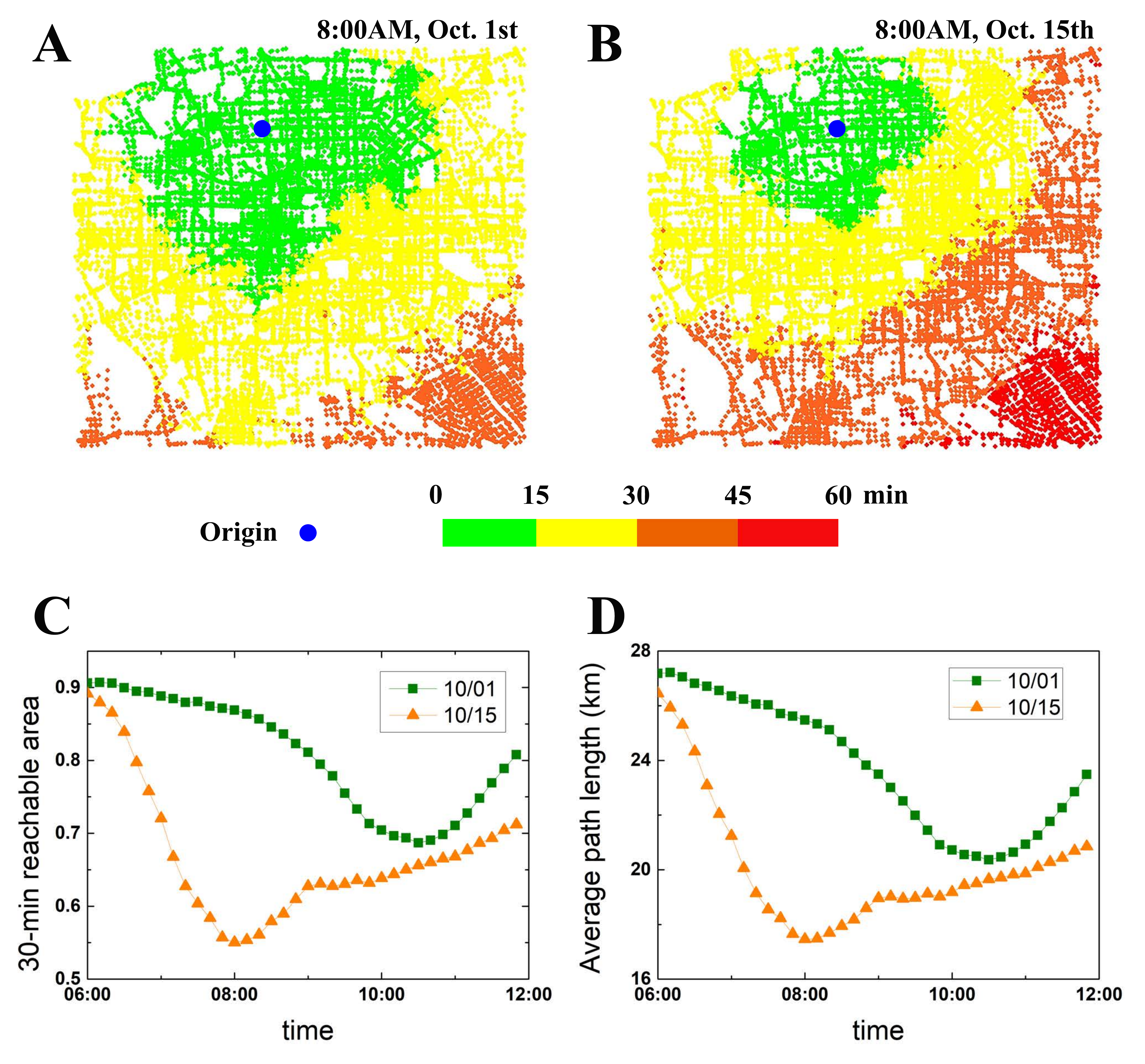}
\caption{\textbf{Reachable area in the dynamic traffic network from a typical site.} Starting from a given site (e.g., Zhichun Road here) in Beijing (marked as a blue circle), the reachable area that one can access within a certain time (i.e., 15 min, 30 min, etc.) at a morning instance on (\emph{A}) a holiday (October 1) and (\emph{B}) a workday (October 15). (\emph{C}) The size of the 30-min reachable area fraction (indicated by the number of reachable nodes divided by the total number of nodes in the road network) for a traveler from a given site in Beijing on the above holiday (squares) and workday (triangles). (\emph{D}) Average path length at the boundary of the 30-min reachable area on the above holiday (squares) and workday (triangles). The results of (\emph{C} and \emph{D}) are averaged by 100 realizations (100 randomly chosen starting sites).}\label{fig:one}
\end{figure*}

The traffic network is considered reliable only when it can convey flows to their destination within a reasonable time. In this sense, congested roads in the network become dysfunctional, and a dynamic network composed of only noncongested roads is available for the driver. The structure of this dynamic network is changing with time, which determines the drivers’ travel time and the whole-network performance. From a starting node, the reachable area (or a distance) for a driver is always constant within a given time for a static network, while this is not the case for a traffic dynamic network \cite{RN59}. Our dataset covers the road network of the Beijing central area, which contains more than 50,000 roads as network links and around 27,000 intersections as network nodes. The velocity dataset covers real-time velocity records of road segments in October 2015, including a representative holiday period in China, the National Day (from October 1 to October 7). Velocity (kilometers per hour) is recorded by GPS devices in floating cars (e.g., taxies and private cars), with resolution of 1 min. We demonstrate in Fig. \ref{fig:one} the reachable area as a function of travel time starting from a typical node near the city center. As time increases, the driver can visit larger areas until reaching the whole network. Between workday and holiday, the reachable area changes significantly for a given travel time. For example, in a holiday shown in Fig. \ref{fig:one}\emph{A}, the driver can reach within 15 min an average distance of around 11 km, and this reachable area will be extended to over 600 km$^2$ (a radius of over 25 km) in 30 min. However, for rush hours during a working day, heavy traffic slows down the velocities and significantly reduces the possible area that one can visit in the same duration (Fig. \ref{fig:one}\emph{B}). One can see from Fig. \ref{fig:one}\emph{C} that the reachable area fraction (indicated by the number of reachable nodes divided by the total number of nodes in the road network) will undergo a major decrease at a rush hour instance on a working day compared with the same time on a holiday. Similar results are shown in Fig. \ref{fig:one}\emph{D}, for the dynamic traffic network during the holiday, drivers can visit a much farther distance in the functional network with higher velocities and less traffic than on workdays.

\begin{figure*}
\centering
\includegraphics[width=17.8cm]{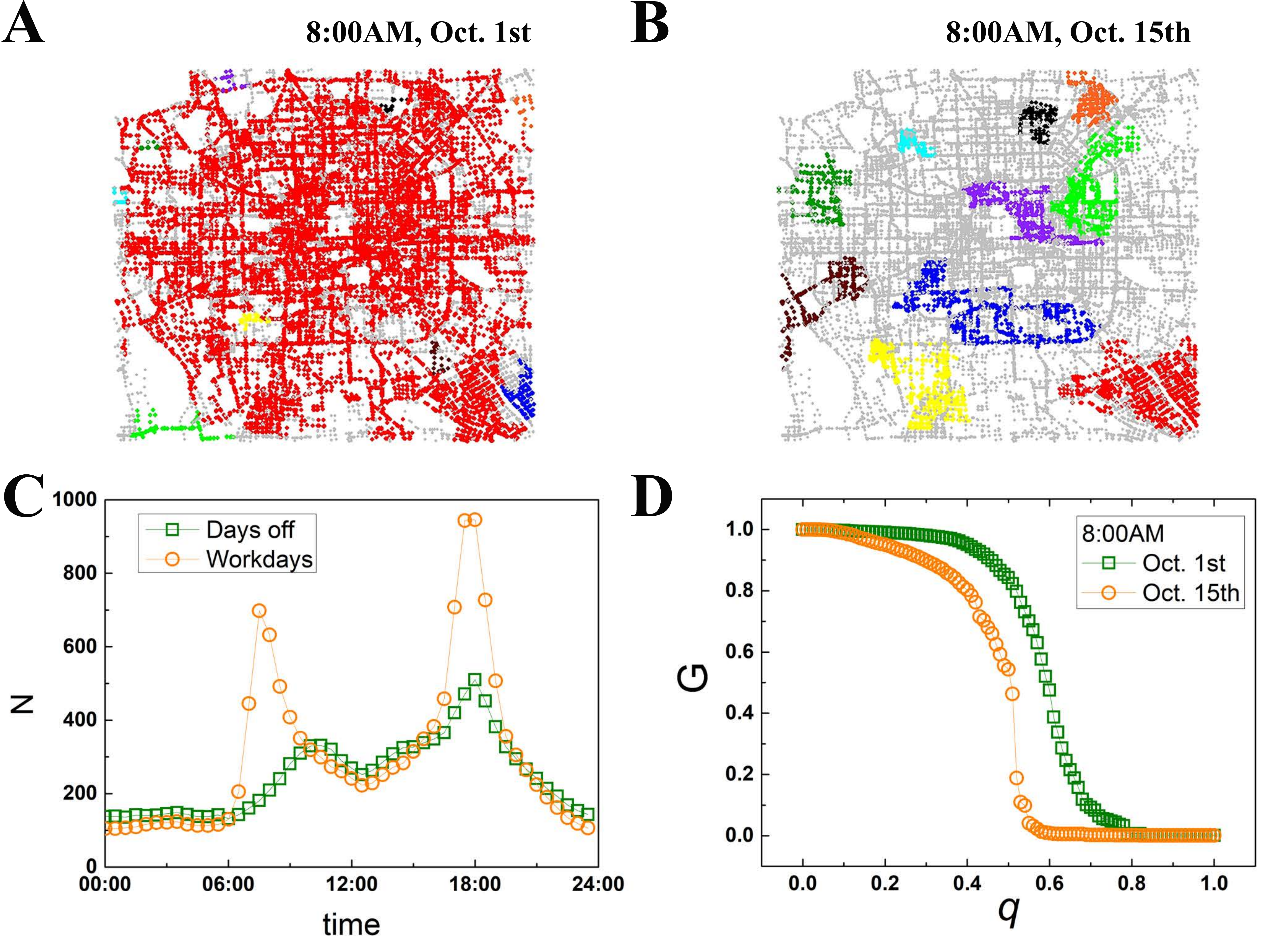}
\caption{\textbf{Robustness of the traffic dynamic network.} (\emph{A} and \emph{B}) Breakdown of traffic clusters under a given value of removed fraction (\emph{q} = 0.55) at the same morning instance during different days. \emph{A} shows a holiday, while \emph{B} shows a working day. (\emph{C}) The number of functional traffic clusters (with high speed) as a function of time. The result is averaged over 12 d off and 17 working days. (\emph{D}) Percolation process [i.e., the giant component of city traffic at a rush hour time on the above holiday (squares) and working day (circles)].}\label{fig:two}
\end{figure*}

Evolution of the dynamic traffic network influences not only the travel efficiency but also, the robustness of this functional network. For holiday, this dynamic network has few link losses (which are below the typical velocity of about 40 km/h) and is close to the original (structural) road network due to the free traffic (Fig. \ref{fig:two}\emph{A}). For rush hours during workdays, highways and other roads become congested due to intensified demand. The large amount of congestion disintegrates the whole dynamic traffic network into isolated functional clusters (connected roads with high speed) (Fig. \ref{fig:two}\emph{B}). As shown in Fig. \ref{fig:two}\emph{C}, the number of isolated functional clusters is increasing during rush hours on both workdays and days off; however, the network is significantly more fragmented on workdays during rush hours, with a much larger number of clusters.

To further investigate the differences in the dynamic organization of the traffic network in the two phases (i.e., rush hours and nonrush hours), we perform a percolation analysis (by calculating the size of the giant component) by deleting the roads with velocity (relative to the maximum velocity) (\emph{SI Appendix}) below \emph{q} from the original road network. In this way, the giant percolation cluster decreases as we increase \emph{q}. The giant component in Fig. \ref{fig:two}\emph{D} shows a phase transition with increasing \emph{q} and becomes fragmented at a critical point (\emph{SI Appendix}, Fig. S1 has more examples). Note that the giant component is decreasing lower and much faster at a morning rush hour instance (8:00 AM) during a workday than during a holiday. This has been further supported with good statistics (\emph{SI Appendix}, Fig. S2) that the giant component during days off and workdays has different distribution at rush hours, raising the possibility of two percolation modes for traffic.

\begin{figure*}
\centering
\includegraphics[width=17.8cm]{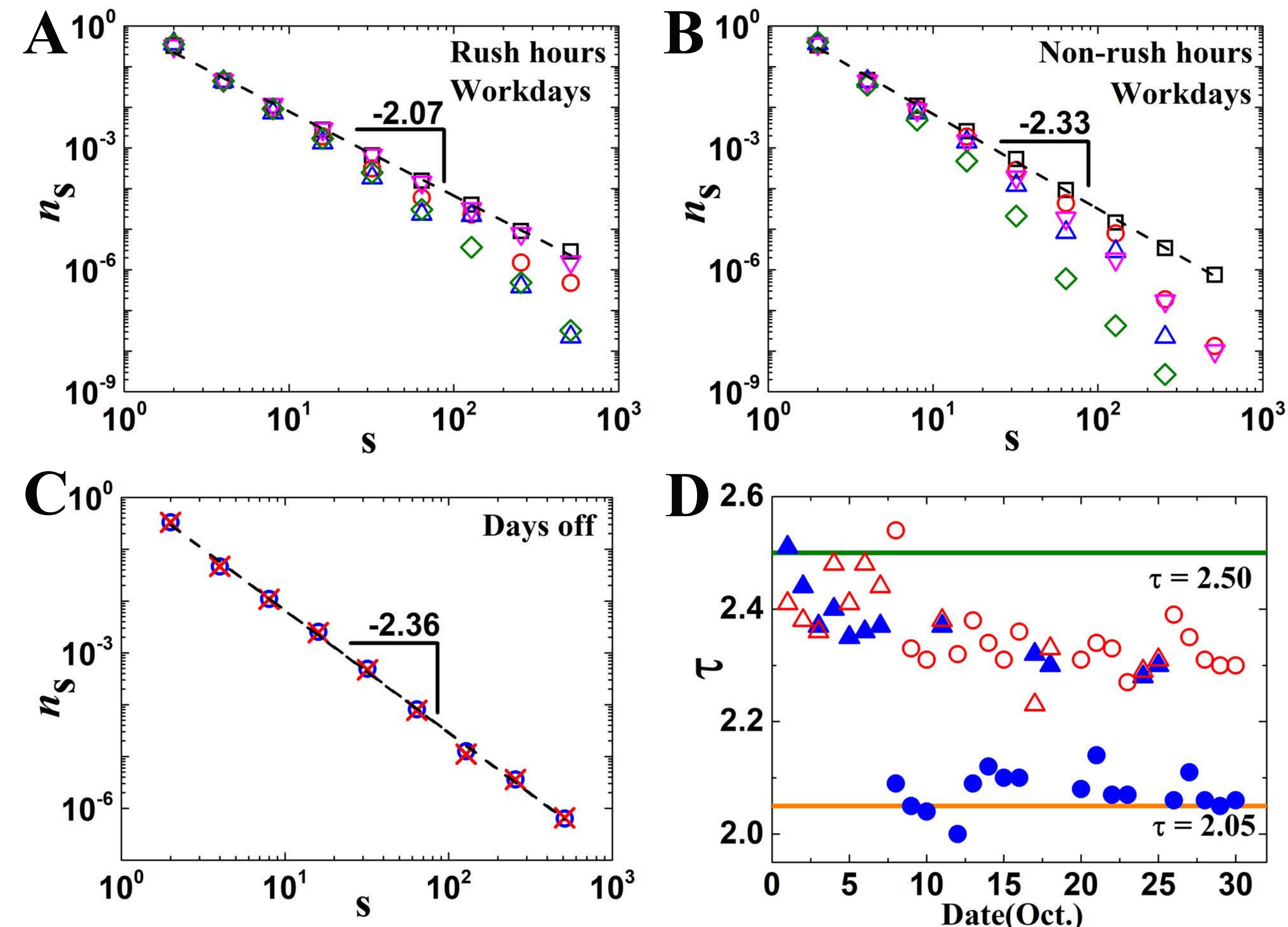}
\caption{\textbf{Percolation critical exponents of cluster size distribution.} (\emph{A} and \emph{B}) Size distribution of traffic flow clusters near criticality during (\emph{A}) rush hours and (\emph{B}) nonrush hours on 17 workdays. Results include size distribution at $q_c$(\emph{t}) (squares), $q_c$(\emph{t})-0.1 (circles), $q_c$(\emph{t})-0.2 (up triangles), $q_c$(\emph{t})+0.1 (down triangles) and $q_c$(\emph{t})+0.2 (diamonds). (\emph{C}) Size distribution of traffic flow clusters at criticality during rush hours (circles) and nonrush hours (crosses) on 12 d off. (\emph{D}) Values of $\tau$ at specific periods of every day, including rush hours on days off (solid triangles), rush hours on workdays (solid circles), nonrush hours on days off (open triangles), and nonrush hours on workdays (open circles). Rush hours here mean 7:30-8:30 AM and 5:30-6:30 PM, while nonrush hours are from 11:00 AM to 1:00 PM every day. The theoretical results of high-dimensional mean field for small world ($\tau$=2.50) and 2D lattice percolation ($\tau$=2.05) are also marked as horizontal lines.}\label{fig:three}
\end{figure*}

Next, we calculate the size distribution of finite clusters of free flows near the critical threshold for the different periods. Fig. \ref{fig:three}\emph{A} and Fig. \ref{fig:three}\emph{B} demonstrates on the workdays the variation of size distribution of finite clusters in the percolation during rush hours and nonrush hours, respectively. At criticality, it is suggested that the size distribution of finite clusters follows a power law \cite{RN1}:

\begin{align*}
n_s \sim s^{-\tau} \numberthis \label{eqn:powerlaw}
\end{align*}

Here \emph{s} is the cluster size, $n_s$ is the ratio between the number of \emph{s}-size clusters and the total number of clusters, and $\tau$ is the corresponding critical percolation exponent. This power law feature can also be seen in day off results in Fig. \ref{fig:three}\emph{C}. As seen in Fig. \ref{fig:three}\emph{A} and Fig. \ref{fig:three}\emph{B}, above the threshold (i.e. $q_c$(\emph{t})+0.1, $q_c$(\emph{t})+0.2), only small clusters composed of high-speed links exist, and the size distribution seems to decay faster than a power law. The change from a power law at criticality to a stronger decay when further from criticality is a sign of $q_c$(\emph{t}) being a critical point \cite{RN1, RN2}. As \emph{q} approaches to the critical threshold $q_c$(\emph{t}), a large cluster (i.e., the giant component) appears, representing the global traffic flow in the traffic network. The same happens when \emph{q} decreases below $q_c$(\emph{t}) (i.e. $q_c$(\emph{t})-0.1, $q_c$(\emph{t})-0.2), the scale of the giant component increases, since more finite clusters are merged into the giant component. Accordingly, the number of finite clusters decreases, leading to a more skewed tail in the size distribution. This behavior further supports the critical percolation hypothesis of traffic flows.

Focusing on the traffic percolation behavior at criticality, we find that the critical exponent of cluster size distribution (denoted by $\tau$) during rush hours on workdays is in general smaller than that during nonrush hours, with values of about 2.07 and 2.33, respectively. However, this difference of critical behavior during different periods does not appear in days off as seen in Fig. \ref{fig:three}\emph{C}. Moreover, the critical exponent during nonrush hours on working days is almost the same as that during days off. These results, therefore, indicate that two modes of percolation critical behaviors exist, suggesting different universality classes for different periods.

Furthermore, we calculate the specific critical exponents of each period for 29 d in 2015 as shown in Fig. \ref{fig:three}\emph{D}. We can see systematic differences in $\tau$ between rush hours on working days and other time periods. Note that, although October 10 was a Saturday, but the exponent $\tau$ is like on a working day. This can be understood by the fact that this specific Saturday was indeed a workday according to the day off compensation policy. It is interesting to note that our results for $\tau$ are generally with values between the theoretical results of high dimension (2.50; e.g., small world or Erdős-Rényi networks) and lattice percolation (2.05; e.g., 2D regular lattice), respectively \cite{RN1, RN2}. During rush hours on working days, the critical exponents are much smaller compared with during other periods, with the value of $\tau$ closer to the limiting case of lattices; on the contrary, during nonrush hours, the values of $\tau$ are significantly and systematically higher. Moreover, some values of $\tau$ during the national holidays are approaching the mean field limit. We show (\emph{SI Appendix}, Fig. S3) that similar findings with different exponents between rush hours and nonrush hours also appear in another large city, Shenzhen. All of these results suggest that the dynamics of city traffic is running at different modes between high-dimensional (mean field) percolation and 2D lattice percolation. Therefore, our results suggest that the value of $\tau$ in real data can be used to classify different traffic modes, which should be managed through different strategies. For cities with only a few highways, like Jinan, we observe (\emph{SI Appendix}, Fig. S5) only the type of percolation transition close to the 2D case.

Thus, the following question can be naturally raised: why do the critical properties behave differently at different periods, although the network structural topology of the roads is the same? A hint to answer the question can come from the knowledge that, for spatially embedded networks \cite{RN40}, the appearance of long-range connections \cite{RN36, RN39} can alter the critical percolation exponents. We suggest here that, for transportation systems, the highways during nonrush hours or days off play the role of effective long-range connections in the traffic network from a dynamic perspective. The highways normally connect distant places and are designed for higher velocities. For a driver, during nonrush hours and days off (in contrast to rush hours), it is usually faster to reach a distant place by highways rather than by other roads. Thus, a plausible hypothesis for the different modes is that, during days off or nonrush hours on working days, there exist more effective long-range connections (high-speed highways) that relax the spatial 2D constraints of the original system. During these periods, the traffic system approaches the traits of small world networks, which behave like a high-dimensional (mean field) system. However, with heavy traffic congestion during rush hours on working days, the highways become congested and are effectively removed from original system.

\begin{figure*}
\centering
\includegraphics[width=17.8cm]{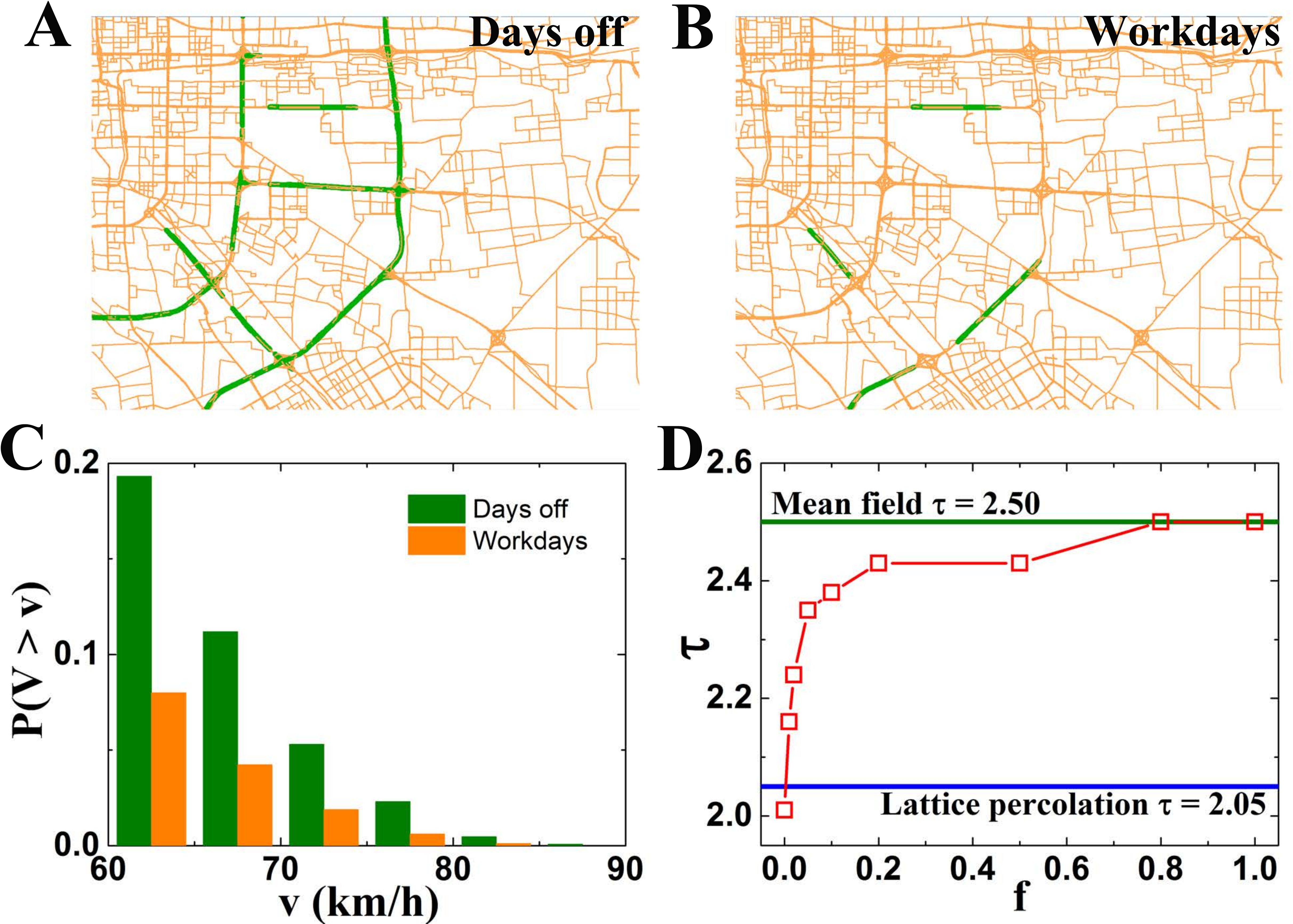}
\caption{\textbf{Effective long-range (high-speed) connections.} (\emph{A} and \emph{B}) Highways with average speed faster than 70 km/h (colored in olive) during rush hours on (\emph{A}) days off and (\emph{B}) workdays. The district shown is a part of the city of Beijing. (\emph{C}) Cumulative velocity distribution of highways on days off (olive) and workdays (orange) in the traffic network. We only focus on velocities faster than 60 km/h. (\emph{D}) Critical exponent $\tau$ as a function of the fraction of rewiring links for percolation in a small world model.}\label{fig:four}
\end{figure*}

To test and demonstrate our assumption, we select the highways of the road network and register their speed during rush hours on workdays and days off. Indeed, we find that the number of highways with high velocities (faster than 70 km/h) is indeed significantly different during the two periods as shown in Fig. \ref{fig:four}\emph{A} and Fig. \ref{fig:four}\emph{B}. During days off, fraction of high-speed highways are found larger in Fig. \ref{fig:four}\emph{C} and effectively form long-range connections, which do not exist during workdays (shown in \emph{SI Appendix}, Fig. S4). We also test our hypothesis by exploring the influence of long-range connections in a lattice percolation model. For that, we apply a link-rewiring model of a 2D small world network \cite{RN36} and study its critical properties. The model network is a lattice with a given fraction \emph{f} of rewired long-range connections, consistent with the fact that the total number of links is kept constant in a city traffic network during different periods. At every step, we randomly choose a node and disconnect one of its links; then, we rewire it to a random node in the whole 2D network. This process continues until the fraction \emph{f} of rewiring links meets the given value.

After constructing the model network, we analyze its percolation properties. The results are shown in Fig. \ref{fig:four}\emph{D}. It suggests that the value of $\tau$ increases with the increasing fraction of rewired links. We can see that the values of $\tau$  are basically changing from 2.05 to 2.50, corresponding to the universality classes of lattice percolation and mean field, respectively. Moreover, $\tau$  changes rapidly as the fraction \emph{f} increases toward 0.2. The larger fraction of links rewired in the lattice, the more similar the network is to the ER network \cite{RN37}. Therefore, our model results support the hypothesis that, during rush hours on workdays, the dynamic percolation of city traffic has a stronger tendency toward 2D lattice percolation, since it is nearly under a zero fraction of long-range connections. Systems with high fractions of long-range connections can exhibit pronounced dynamic effects, which will not occur in systems with short-range interactions only. The existence of long-range connections can result in quasistationary states with slow relaxation toward steady states in driven dynamics \cite{RN42}. As seen here (Fig. \ref{fig:four}\emph{D}) a small fraction of long-range connection can affect the critical percolation exponents, resulting in changes of the system university class \cite{RN14}. In analogy, for example, studies have been conducted on the behavior of the Ising model in a small world network \cite{RN45}, and they find that the introduction of long-range connections (reflected by rewiring probability \emph{f} of network links) results in changing the system universality class from a pure low-dimensional system for $f = 0$ to mean-field like region for $f > 0$. For city traffic systems, this change is self-organized with the variant high-speed roads that represent effective long-range connections (results are in \emph{SI Appendix}, Fig. S6), leading to the mode switching. Thus, by adjusting effective high-speed connections, one may change the system to the desired universality class of city traffic (like during nonrush hours or days off). Note that our results here are obtained for percolation, neglecting the correlation presence in the congestion scenario of real traffic. However, it is suggested \cite{RN60} that several percolation critical exponents, including exponent $\tau$ of cluster size distribution, almost do not change with the introduction of correlations.

\subsection*{Conclusion}

While the growth of urbanization is expected to strain urban infrastructure across all transportation modes, different smart city frameworks have been proposed to combat this challenge. One of the core tasks for city resurgence is to build a resilient transportation system that can adapt to various perturbations and recover from major congestions efficiently. On this task, one fundamental question is how urban traffic will break when approaching its critical point. City traffic is generally a spatial–temporal system \cite{RN56}, hence, it is essential to focus not only on the static structure of a road network but also, on the dynamic organization of traffic in which demand changes from time to time and place to place during a day. Here, we study the percolation transition classification of city traffic dynamics using high-resolution real-time traffic data. By analyzing the cluster size distribution of city traffic, we find that the disintegration transition of urban traffic can be characterized by two sets of percolation critical exponents. During workday rush hours, the critical percolation exponent is close to that of 2D lattice percolation, while during other periods, it is closer to the high-dimensional small world systems.

Our findings suggest that a key point affecting the critical exponent of traffic dynamics is the fraction of the effective long-range connections represented by the connected roads of high speed. Thus, with the aid of dynamic traffic management methods, it may be possible to shift the system to the desired critical universality class by adjusting the amount of effective long-range connections. For demonstration, if appropriate actions can keep the fraction of effective long-range connections above 0.1, we may expect percolation critical exponents to be around 2.4, which represents a universality class of significantly better and more smoothly fragmented global traffic. While the effective long-range connections are the products of structural urban highways, their management also requires much effort on the design of corresponding management strategy along with signal control \cite{RN54} or road pricing \cite{RN55}. In this sense, our study may be useful for modeling and designing traffic resilience in the realization of the smart city. Furthermore, our approach is general and can be applied to other critical infrastructures that are aiming to transfer flow of different types.\\

\newpage

\newpage

\newpage

\newpage

\bibliography{References}

\begin{thebibliography}{10}

\bibitem{RN11}
Dorogovtsev SN, Goltsev AV, Mendes JF (2008) Critical phenomena in complex
  networks.
\newblock {\em Reviews of Modern Physics} 80(4):1275--1335.

\bibitem{RN1}
Bunde A, Havlin S (1991) {\em Fractals and disordered systems}.
\newblock (Springer, Berlin).

\bibitem{RN2}
Stauffer D, Aharony A (1994) {\em Introduction to percolation theory}.
\newblock (CRC press, London).

\bibitem{RN4}
Pastor-Satorras R, Vespignani A (2001) Epidemic spreading in scale-free
  networks.
\newblock {\em Physical Review Letters} 86(14):3200--3203.

\bibitem{RN6}
Meloni S, Arenas A, Moreno Y (2009) Traffic-driven epidemic spreading in
  finite-size scale-free networks.
\newblock {\em Proceedings of the National Academy of Sciences}
  106(40):16897--16902.

\bibitem{RN8}
Stanley HE, Stauffer D, Kertesz J, Herrmann HJ (1987) Dynamics of spreading
  phenomena in two-dimensional {Ising} models.
\newblock {\em Physical Review Letters} 59(20):2326--2328.

\bibitem{RN17}
Chowdhury D, Santen L, Schadschneider A (2000) Statistical physics of vehicular
  traffic and some related systems.
\newblock {\em Physics Reports} 329(4):199--329.

\bibitem{RN18}
Helbing D (2001) Traffic and related self-driven many-particle systems.
\newblock {\em Reviews of Modern Physics} 73(4):1067--1141.

\bibitem{RN19}
Kerner BS (2004) {\em The physics of traffic}.
\newblock (Springer, New York).

\bibitem{RN20}
Daganzo CF (2007) Urban gridlock: Macroscopic modeling and mitigation
  approaches.
\newblock {\em Transportation Research Part B: Methodological} 41(1):49--62.

\bibitem{RN33}
Li D, et~al. (2015) Percolation transition in dynamical traffic network with
  evolving critical bottlenecks.
\newblock {\em Proceedings of the National Academy of Sciences}
  112(3):669--672.

\bibitem{RN21}
Lighthill M, Whitham G (1955) On kinematic waves. i. flood movement in long
  rivers in {\em Proceedings of the Royal Society of London A: Mathematical,
  Physical and Engineering Sciences}.
\newblock (The Royal Society), Vol.{} 229, pp. 281--316.

\bibitem{RN22}
Prigogine I, Herman R (1971) {\em Kinetic theory of vehicular traffic}.
\newblock (Elsevier, New York).

\bibitem{RN23}
Newell GF (1993) A simplified theory of kinematic waves in highway traffic,
  part i: General theory.
\newblock {\em Transportation Research Part B: Methodological} 27(4):281--287.

\bibitem{RN25}
Kerner BS (1998) Experimental features of self-organization in traffic flow.
\newblock {\em Physical Review Letters} 81(17):3797--3800.

\bibitem{RN26}
Nagel K, Schreckenberg M (1992) A cellular automaton model for freeway traffic.
\newblock {\em Journal de physique I} 2(12):2221--2229.

\bibitem{RN27}
Helbing D, Huberman BA (1998) Coherent moving states in highway traffic.
\newblock {\em Nature} 396(6713):738--740.

\bibitem{RN57}
Jiang R, et~al. (2015) On some experimental features of car-following behavior
  and how to model them.
\newblock {\em Transportation Research Part B: Methodological} 80:338--354.

\bibitem{RN12}
Cohen R, Havlin S (2010) {\em Complex networks: structure, stability and
  function}.
\newblock (Cambridge Univ Press, Cambridge, UK).

\bibitem{RN13}
Barzel B, Barabási AL (2013) Universality in network dynamics.
\newblock {\em Nature Physics} 9:673--681.

\bibitem{RN51}
Geroliminis N, Sun J (2011) Properties of a well-defined macroscopic
  fundamental diagram for urban traffic.
\newblock {\em Transportation Research Part B: Methodological} 45(3):605--617.

\bibitem{RN58}
Majdandzic A, et~al. (2014) Spontaneous recovery in dynamical networks.
\newblock {\em Nature Physics} 10(1):34--38.

\bibitem{RN10}
Gao J, Barzel B, Barabási AL (2016) Universal resilience patterns in complex
  networks.
\newblock {\em Nature} 530(7590):307--312.

\bibitem{RN59}
Weiss D, et~al. (2018) A global map of travel time to cities to assess
  inequalities in accessibility in 2015.
\newblock {\em Nature} 553(7688):333--336.

\bibitem{RN40}
Barthélemy M (2011) Spatial networks.
\newblock {\em Physics Reports} 499(1):1--101.

\bibitem{RN36}
Watts DJ, Strogatz SH (1998) Collective dynamics of ‘small-world’networks.
\newblock {\em Nature} 393(6684):440--442.

\bibitem{RN39}
Redner S (2008) Networks: teasing out the missing links.
\newblock {\em Nature} 453(7191):47--48.

\bibitem{RN37}
Barthélémy M, Amaral LAN (1999) Small-world networks: Evidence for a
  crossover picture.
\newblock {\em Physical Review Letters} 82(15):3180--3183.

\bibitem{RN42}
Campa A, Dauxois T, Ruffo S (2009) Statistical mechanics and dynamics of
  solvable models with long-range interactions.
\newblock {\em Physics Reports} 480(3):57--159.

\bibitem{RN14}
Daqing L, Kosmidis K, Bunde A, Havlin S (2011) Dimension of spatially embedded
  networks.
\newblock {\em Nature Physics} 7(6):481--484.

\bibitem{RN45}
Herrero CP (2002) Ising model in small-world networks.
\newblock {\em Physical Review E} 65(6):066110.

\bibitem{RN60}
Prakash S, Havlin S, Schwartz M, Stanley HE (1992) Structural and dynamical
  properties of long-range correlated percolation.
\newblock {\em Physical Review A} 46(4):R1724--R1727.

\bibitem{RN56}
Brockmann D, Helbing D (2013) The hidden geometry of complex, network-driven
  contagion phenomena.
\newblock {\em Science} 342(6164):1337--1342.

\bibitem{RN54}
Papageorgiou M, Diakaki C, Dinopoulou V, Kotsialos A, Wang Y (2003) Review of
  road traffic control strategies.
\newblock {\em Proceedings of the IEEE} 91(12):2043--2067.

\bibitem{RN55}
Yang H, Huang HJ (2005) {\em Mathematical and economic theory of road pricing}.
\newblock (Elsevier, Oxford).

\end{thebibliography}

\end{document}